\documentclass[10pt,twocolumn,conference]{IEEEtran}
\usepackage{amsfonts}
\usepackage{graphicx}
\usepackage{latexsym}
\usepackage{amsmath}
\usepackage{cite}
\usepackage{graphicx}
\usepackage{subfigure}

\newcommand{\define}{\stackrel{\Delta}{=}}

\begin{document}

\title{Precoded Turbo Equalizer for Power Line Communication Systems}
\author{Kai Xie, \ and \ Jing Li (Tiffany) \\
Electrical and Computer Engineering Dept, Lehigh University,
Bethlehem, 18015, USA\\ Emails: kax205@lehigh.edu,
jingli@ece.lehigh.edu  }

\maketitle

\footnotetext{Supported by NSF under Grants No. CCF-0635199,  
CCF-0829888 and CMMI-0928092.}

\begin{abstract}
Power line communication continues to draw increasing interest by promising a wide range of applications including cost-free last-mile communication solution.
However, signal transmitted through the power lines deteriorates
badly due to the presence of severe inter-symbol interference (ISI) and
harsh random pulse noise. This work proposes a new precoded turbo equalization
scheme specifically designed for the PLC channels. By introducing useful precoding to reshape ISI, optimizing  maximum {\it a posteriori} (MAP) detection to address the  non-Gaussian pulse noise, and performing soft iterative decision refinement, the new equalizer demonstrates a gain significantly better than the existing turbo equalizers.
% on PLC in three aspects: performance gain by introducing
%precoder, performance gain from the optimized MAP algorithm, and a
%design guide for precoders.
\end{abstract}

%\begin{keywords}
%power line communication, inter-symbol interference, extrinsi

%%%%%%%%%%%%%%%%%%%%%%%%%%%%%%%%%%%%%%%%%%%%%%%%%%%%%%%
%%%%%%%%%%%%%%%%%%%%%%%%%%%%%%%%%%%%%%%%%%%%%%%%%%%%%%%
\section{Introduction}
\label{sec:introduction}

The inception of power line communication (PLC) has lead to many useful applications from remote voltage monitoring, to  meter readings of power line systems, %and then extended to many important applications:
 broadband Internet access, and more recently, indoor wired local area networks.
%A significant advantage of PLC is its ubiquity, which makes
The ubiquity of power line makes it a promising candidate for providing an cost-free last-mile communication solution.
%infrastructure.
%Therefore, it is nearly perfect for the last mile
%communication solution, and has drawn a considerable attention from
%researchers.

%Several standards and practical systems have been proposed and are
%employed for various purposes. Different types of PLC system
%employing different frequency bands and voltages suffer from various
%signal transmission characteristics. The electric power lines are
%categorized by bandwidths or voltages. By bandwidths, they are
%grouped into broadband with high frequency($1M-30M$) and narrow band
%with low frequency($<1M$). By voltages, they are classified into the
%medium ($1-100 kV$) and low (smaller than $1$ kV) voltage networks,
%while high-voltage lines are usually not used for data transmission
%due to excessive noise.

However, the power line system was not originally designed to transmit (data)
signal. Data transmission over power lines generally suffers from harsh random pulse noise and serious multipath fading caused by the impedance mismatch, as branches of the power line network reflect signal back and create several signal
paths from a transmitter to a receiver. Equalization techniques are
therefore actively exploited to cope with  the severe multipath fading, or, the
inter-symbol interference (ISI) effect, in PLC channels.

Traditional channel equalizers, including transversal equalizers, decision feedback equalizers (DFE) and maximum-likelihood sequence equalizers
(MLSE), are optimized under
Gaussian noise. In PLC systems, nonlinear equalizers such as radial basis
function (RBF) networks and fuzzy equalizers \cite{bib:Sail99,bib:Ribe03,bib:Ribe07,bib:Wong07} are also favorable choices, and several studies have demonstrated their good performances in the presence of serious uncertainty posed by the pulse noise.

Another important branch of equalization technique combines channel coding and channel equalization, known as {\it turbo equalization}. Incepted in 1995 \cite{bib:Doui95}, it borrows principles from
turbo codes, % \cite{bib:Berr96},
and is capable of achieving a remarkable {\it interleaving gain} by  jointly performing maximum {\it a posteriori} (MAP) detection \cite{bib:BCJR} or near-MAP detection for the ISI channel (inner code) and soft-decoding the channel code (outer code) through an iterative process.
The considerable success of turbo equalizers in radio frequency (RF) wireless communication systems has also promoted a serious investigation of its applicability and performance in PLC. Specifically, a pioneering work in \cite{bib:Chuah07} proposed to complement the conventional turbo equalizer with a front-end myriad filter (MMyF)  for efficient baseband filtering in impulsive channels. The use of MMyF was shown to bring encouraging gains than otherwise, but the capacity
potential of turbo equalization was still not fully exploited in the system discussed in \cite{bib:Chuah07}. The reason is
two-fold: 1) with the pulse noise processed outside the turbo
equalizer, the turbo equalizer was not optimized with the pulse noise taken into consideration; and 2) the PLC ISI channel was not precoded.
The latter is particularly pitiful, because from the coding theory, we know that the interleaving gain of a serially concatenated turbo system is attainable only when the inner code is {\it recursive}. In the case of turbo equalization, as  an ISI channel (e.g. the PLC channel) acts as an inner code, it  is by nature \emph{non-recursive}, and must therefore be precoded in order to appear {\it recursive}.

This work proposes a new turbo equalizer scheme to overcome these
two defects. In the new scheme, a precoder is carefully applied to re-structure the PLC channel to a recursive ISI channel, and the conventional MAP detector is re-designed and tailored to the precoded ISI channel and especially to the
presence of random pulse noise. Additionally, we propose a modified
extrinsic information transfer chart (EXIT) approach to simplify the
design of precoder. Simulation results demonstrate that the
new turbo equalizer scheme noticeably outperforms the existing ones and that the modified EXIT chart provides a useful tool for selecting appropriate precoders.

\section{Channel model}
\label{sec:channel model}

Data communication over power lines is subject to a number of
destructive interferences and impairments including signal attenuation caused by
cable loss, ISI caused by multipath propagation, additive white Gaussian
noise, and intermittent strong pulse noise.
%which in turn leads to  a multi-path fading channel with additive noise.

Consider feeding a binary data sequence $w_i$ into a transmit filter with pulse
shape  $g(t)$. The shaped signal $x(t)$ takes the form of
%------------------
\begin{equation}\label{equ:transmited signal}
    x(t) = \sum_{i=1}^{\infty}w_ig(t-iT).
\end{equation}
%-----------------
After the signal passes through the PLC channel with frequency-domain impulse
response $H_C(f)$, time-domain impulse response $h_c(t)$ and additive noise $z(t)$, the receiver gets
%-----------------
\begin{equation}\label{equ:channel model in time domain}
    r(t) = \sum_{i=1}^{\infty}w_ich(t-iT)+z(t)
\end{equation}
%-----------------
where
%-----------------
\begin{equation}\label{equ:time0-domain channel coefficients}
    ch(t) = \int_{-\infty}^{\infty} g(\tau)h_c(t-\tau)d\tau.
\end{equation}
%-----------------
%

This PLC channel model can also be expressed in a discrete form:
%-----------------
\begin{equation}\label{equ: descrete channel model in time-domain}
    y[n] = \sum_{k=1}^{L_h-1} h[k] w[n-k] + Noise[n],
\end{equation}
%-----------------
where $h[k]$ is the sampled sequence of the channel response $ch[t]$,
$L_h$ is the number of (multi) paths, $Noise[n]$ is the sampled additive noise and $y[n]$ is the discrete-form received signal at the $n$-th time slot.

%\subsection{Inter symbol interference (ISI) channel model} \label{sec:ISI channel model}

The multi-path fading of a PLC channel results from the tree-like
topology of the PLC network with multiple branches. These branches have
different lengths and are loaded with various impedance. The
transmitted wave suffers from reflections at every impedance
mismatch point caused by the difference between characteristic
impedance of the cables. Consequently, instead of propagating along a
single path, the signal reflects along different branches and forms a
multi-path channel. Generally, the signal attenuation along the PLC channel
increases with the distance.

The most widely known and cited frequency-domain PLC channel model
\cite{bib:Zimm02a}  approximates the overall channel response $H_c(f)$ by emphasizing
 the most dominant set of paths that exist over the frequency range of 500 kHz to 20 MHz:
%---------------------
\begin{equation}\label{equ:frequence channel model}
H_C(f) = \sum_{i=1}^{L_f} \xi_i e^{-(a_0+a_1 f^\kappa)d_i} e^{-i2\pi
f(d_i/v_p)}
\end{equation}
%---------------------
where $L_f$ is the total number of paths, whose typical value ranges  from $3$ to $5$. The attenuation of the $i$-th
path increases  exponentially with the distance $d_i$ where the attenuation power factor is  determined by some parameters $\{a_0, a_1\}$ and
$\kappa$, where $\kappa$ is typically in the interval of ($0.2$, $1$).
All the $L_f$ paths accumulate like a weighted sum with weight $\xi_i$ for the $i$-th path.  The last term $e^{-i2\pi f(d_i/v_p)}$ represents the propagation delay with the velocity of propagation parameter $v_p$, which can be calculated by
%---------------------
\begin{equation}\label{equ:velocity of propagation}
    v_p = \frac{C_0}{\sqrt{\varepsilon_r}},
\end{equation}
%--------------------
where $C_0$ is the speed of light, and $\varepsilon_r$ represents
the dielectric constant of the insulating material.

%This ISI channel model \cite{bib:Zimm02a} has shown excellent agreement with the measured values.

Besides ISI, a PLC channel also surfers from a harsh non-Gaussian
additive noise, which is commonly assumed to be a composition of five sources \cite{bib:Zimm02b}\cite{bib:Dost01}: background noise,
narrow band interference noise coming from the surrounding radio signals, and pulse noise caused by the fundamental component of the power system, by the switching power supplies, and by the random switching transients in the power line network.
%--------------------
%\begin{equation}\label{equ: noise model1}
%    Noise = N_{bkgr}+ N_{nb} + N_{pa} + N_{ps} + N_{imp},
%\end{equation}
%--------------------
%where $N_{bkgr}$ is the background noise, $N_{nb}(n)$ is a narrow
%band interference noise coming from the surrounding radio signals, and
%$N_{pa}$, $N_{ps}$ and $N_{imp}$ are all pulse noise. $N_{ps}(n)$ is
%periodic pulse noise composite caused by the fundamental component
%of the power system. $N_{pa}(n)$ is a periodical pulse noise that is
%asynchronous to the fundamental component of the power system and caused
%by switching power supplies. $N_{imp}(n)$ is also an asynchronous
%pulse noise caused by random switching transients in the power line
%network.
%The cumulative effect of these noisee sources is a combination of  additive white Gaussian noise and random impulses.
 The prevailing statistical model to characterize the cumulative effect of all this additive noise is  a two-term Gaussian mixture model \cite{bib:Dai03} \cite{bib:Wong07}:
%------------------
\begin{equation}\label{equ: noise model2}
    Noise = (1-\varepsilon){\cal N}(0,\sigma^2) + \varepsilon
   {\cal  N}(0,K\sigma^2),
\end{equation}
%------------------
where $\varepsilon$ represents the probability of impulses, and
${\cal N}(0,\sigma^2)$ and ${\cal N}(0,K\sigma^2)$ are Gaussian
distributions with zero mean and variances of $\sigma^2$ and
$K\sigma^2$, respectively. In the model of (\ref{equ: noise
model2}),  the channel is for the most time dominated by the
background noise and the narrowband interference, which are
collectively represented by an Gaussian distribution ${\cal
N}(0,\sigma^2)$; and with a small probability of $\varepsilon$, it
is overwhelmed by  the pulse noise, which is  modeled by another
independent Gaussian distribution ${\cal N}(0,K\sigma^2)$ with a
much larger variance.

%This two-term model can also be extended to a multi-term Gaussian
%mixture \cite{bib:Umeh06}, such that
%------------------
%\begin{equation}\label{equ: noise model3}
%    Noise = \sum_i^D P^{ini}_i {\cal N}(0,\sigma_i^2),
%\end{equation}
%%------------------
%where $\sum_i P^{ini}_i = 1$ and $D$ is the total number of
%terms. Testing and verification of this extended additive noise
%model for PLC took place in Japan with three different configurations,
%and it is shown that the mixed Gaussian noise model (\ref{equ: noise model3})
%matches with the practical channel characteristics very well. Hence, in this work, we  adopt this multi-term Gaussian mixture model as our additive noise model.
%{\bf [Did you actually simulate the two-term model or the more-than-two-term model? If it is the former, then you should probably delete this paragraph.]}

%%%%%%%%%%%%%%%%%%%%%%%%%%%%%%%%%%%%%%%%%%%%%%%%%%%%%%%
%%%%%%%%%%%%%%%%%%%%%%%%%%%%%%%%%%%%%%%%%%%%%%%%%%%%%%%
\section{Design of Turbo Equalizers for PLC Channels}
\label{sec:turbo equalizer}

\subsection{System Model}
The remarkable performance of turbo equalizers in
combating ISI has been demonstrated in a rich variety of wireless systems and applications \cite{bib:Doui95}. At the
transmitter side, by serially connecting the ISI channel and an
error correction code (ECC) through an interleaver, turbo equalizer can
treat the ISI channel as the inner code and the ECC as the outer code of a serially concatenated system.
At the receiver side, both of the component codes are softly-decoded and the soft extrinsic
information is exchanged back and forth between them to iteratively refine the decision (i.e. iterative processing instead of the conventional sequential one-way processing).

%Different from SCCC, the turbo equalizer replaces the binary
%recursive convolutional inner code with a real-valued non-recursive
%convolutional channel. According to \cite{bib:Bene96}, the recursive
%convolutional code can map a low-weight input sequence to a
%high-weight codeword, while the non-recursive code can only generate
%a low-weight codeword for the low-weight input.
It is widely recognized that the inner code of an interleaved serially concatenated system must be recursive, in order to effectively achieve the spectrum thinning effect and hence attain the so-called interleaving gain \cite{bib:Bene96}.
% convolutional code is very important for turbo structure
%and will lead to a better weight spectrum. The gain from the
%recursive structure is also named as interleaving gain promised by a
%turbo-like concatenated system.
 In a PLC system, the interleaving gain becomes particularly important and desirable, largely due to the long delay spread of the PLC channel. However, an ISI channel is by nature non-recursive, that is, the output from the channel is the result of a non-recursive convolution between the input signal and the ISI channel. Hence, to obtain the interleaving gain, it is necessary to re-shape the channel by adding a rate-1 recursive precoder before the ISI
channel.

%Besides the inter symbol interference, PLC channels also suffer from
%the interference of pulse noise. To mitigate the pulse noise,
%\cite{bib:Chuah07} proposed a structure by adding a matched myriad
%filter before the turbo equalizer. However, since the matched filter
%and the equalizer performed separately and there is no soft
%information exchanged between them, the structure is not globally
%optimized. The iterative searching algorithm of matched myriad
%filter is computational costy. To improve the performance, we
%proposed a new MAP equalizer algorithm which is capable of
%processing the pulse noise and the ISI at the same time. To further
%improve the performance, we also add a precoder after the
%interleaver to make the channel recursive.

\begin{figure}[h] \centering
\includegraphics[width=3.5in]{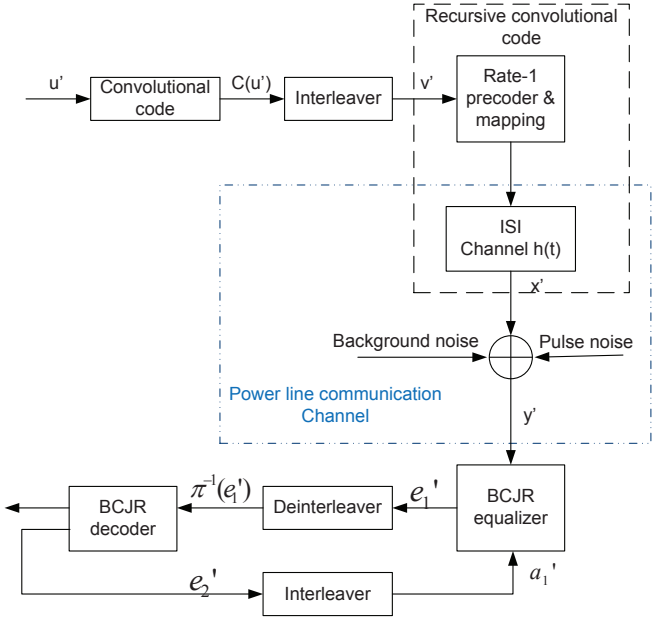}
\caption{System diagram of precoded turbo equalizer.}
\label{fig:system block of precoded turbo equalizer}
\end{figure}

Figure \ref{fig:system block of precoded turbo equalizer} shows the
system model. A binary sequence $\vec{u}$ with length of $N$ is
encoded by an outer code, which, in this specific example, is a recursive systematic convolutional (RSC) code $C1$. The coded sequence from the outer code $\vec{v}$ is then interleaved, and subsequently passed through a rate-1
recursive convolutional code $C2$ (the precoder), and finally sent through the PLC channel. The combination of the ISI channel and the recursive precoder acts like a recursive convolutional inner code, whose output sequence is denoted by $\vec{x}$. At the receiver side, the received sequence $\vec{y}$ is fed into an
iterative decoder consisting of two sub-decoders, the maximum {\it a posteriori} (MAP) equalizer and the BCJR decoder which are matched, respectively, to the precoded ISI channel (the inner code) and the outer convolutional code. Soft extrinsic information from the these sub-decoders, $\vec{e}_1$ and $\vec{e}_2$, are exchanged to iteratively refine the detection and decoding decisions.

%--------------------------------
%---------------------------------

\subsection{Matching MAP Equalizer to Pulse Noise}
\label{sec:MAP}
Although  not having been investigated in the context of PLC before, the precoded equalizer configuration in Fig. \ref{fig:system block of precoded turbo equalizer}) has been widely exploited and found great success in the radio frequency (RF) wireless systems. However, the special characteristics of the PLC channel (e.g. strong pulse noise) make a direct adaption of this configuration suboptimal. To fully harness the gain promised by the theory requires the addressing of two technical issues: how to get the precoder done right and how to get the MAP equalizer done right. Below we will discuss the second issue, and will leave the first to the next section.

%However, a PLC channel differs from an RF channel in that it has severe random pulse noise, which makes the conventional MAP equalizer, .

The conventional MAP equalizer, be it for precoded or non-precoded ISI channel,  is derived with the assumption that the additive noise is always Gaussian and white \cite{bib:BCJR}. When the noise is not Gaussian, such as the case in PLC, a channel mis-match results, causing the conventional MAP to degrade.
% perform suboptimally.

To cope with the pulse noise, the MAP equalization algorithm
must be re-derived. In the MAP equalizer,  the log-likelihood ratio (LLR) associated with the bit $v_k$ is computed as \cite{bib:BCJR}
%------------------------------
\begin{align}
  L(v_k) &= log\frac{P(v_k = 1|\vec{y})}{P(v_k = 0|\vec{y})}
%   &=& log\frac{\displaystyle\sum_{(s_{k-1},s_k) \in S^1}P(s_{k-1},s_k|\vec{y})}{\displaystyle\sum_{(s_{k-1},s_k) \in S^0}P(s_{k-1},s_k|\vec{y})} \\
   = log\frac{\displaystyle\sum_{(s_{k-1},s_k) \in S^1}P(s_{k-1},s_k,\vec{y})}{\displaystyle\sum_{(s_{k-1},s_k) \in S^0}P(s_{k-1},s_k,\vec{y})} \nonumber\\
%   &=& log\frac{\displaystyle\sum_{(s_{k-1},s_k) \in S^1} P(s_{k-1}, \vec{y}_0^{k-1})P(s_{k-1},s_k, y_k)P(s_k,\vec{y}_{k+1}^{N-1})}{\displaystyle\sum_{(s_{k-1},s_k) \in S^1} P(s_{k-1},
%   \vec{y}_0^{k-1})P(s_{k-1},s_k,y_k)P(s_k,\vec{y}_{k+1}^{N-1})}\\
   &= log\frac{\displaystyle\sum_{(s_{k-1},s_k) \in S^1} \alpha(s_{k-1})\gamma(s_{k-1},s_k)\beta(s_k)}{\displaystyle\sum_{(s_{k-1},s_k) \in S^0} \alpha(s_{k-1})\gamma(s_{k-1},s_k)\beta(s_k)},
\label{equ:LLR_BCJR}
\end{align}
%-----------------------------
where $s_k$ represents the state of the trellis at the time index of
$k$, $S^1$ and $S^0$ denote the sets of branches $\{s_{k-1},s_k\}$
corresponding to $v_k = 1$ and $v_k = 0$, respectively, and
$\alpha_k$, $\beta_k$ and $\gamma_k$ are the forward path metric,
the backward path metric, and the branch metric, respectively.

%\begin{equation}\label{equ: gamma of BCJR}
%    \gamma_k(s_{k}=s, s_{k-1}=s') = P(y_k, s_{k}=s|s_{k-1=s'}).
%\end{equation}
%
%If the Gaussian channel with noise of $N(0, sigma^2)$ applied, the
%$\gamma$ is calculated as
%
%\begin{equation}\label{equ: gamma for Gaussian noise}
%    \gamma_k(s_k=s, s_{k-1}=s')  =
%    e^{\frac{1}{2}a_ku_k+\frac{L_c}{2}y_kx_k-\frac{L_c}{4} x_k^2},
%\end{equation}
%where $a_k$ is the extrinsic information from another sub-decoder
%and $L_c$ is a parameter determined by the noise level $L_c =
%2/\sigma^2$.
%
%Instead of the single Gaussian noise model, the additive noise in
%PLC channel is a mixture of the Gaussian noises. Different symbol
%may enjoy different variances. We will perform corresponding
%modification on the calculation of $\gamma$.
%--------------------------------
The branch metric $\gamma_k$ must be modified to reflect the specific channel condition. When the noise in the PLC is specified as in (\ref{equ: noise model2}), $\gamma_k$ takes the form of:
\begin{eqnarray}
  &&\gamma_k (s_k=s, s_{k-1}=s')\nonumber \\
  &=& P(y_k, s_k|s_{k-1})
         = P(y_k|x_k)P(u_k) \nonumber \\
         &=& P(u_k) \sum_i P_i(\sigma_i|x_k) P(y_k|x_k,\sigma_i) \nonumber \\
         &=& P(u_k) \sum_i P^{ini}_i(\sigma_i)
         \frac{1}{\sigma_i}e^{\frac{-(x_k-y_k)^2}{\sigma_i^2}}.
\label{equ:gamma}
\end{eqnarray}
%--------------------------------

The forward and backward path metrics capture the PLC channel characteristics through the branch metric $\gamma_k$. Their mathematical forms
 follow a recursive way of computing \cite{bib:BCJR}:
\begin{align}\label{equ: alpha of BCJR}
  \alpha_k(s) &= \!P(s_{k-1}=s, \vec{y}_0^{k-1})
\!   = \!\sum_{s'}\alpha_{k-1}(s')\gamma(s',s),\\
\label{equ: beta of BCJR}
  \beta_k(s) & =\! P(\vec{y}_{k+1}^{N-1}|s_k=s)
\!   =\! \sum_{s'}\beta_{k+1}(s')\gamma(s,s').
\end{align}
%-------------------

Inserting (\ref{equ:gamma})-(\ref{equ: beta of BCJR}) in (\ref{equ:LLR_BCJR}) leads to a modified soft-input soft-output MAP algorithm tailored for the specific channel model of the PLC.

%%%%%%%%%%%%%%%%%%%%%%%%%%%%%%%%%%%%%%%%%%%%%%%%%%%%%%%
%%%%%%%%%%%%%%%%%%%%%%%%%%%%%%%%%%%%%%%%%%%%%%%%%%%%%%%
\section{Precoder Design through Exit Analysis}
\label{sec:exit chart analysis of turbo equalizer}

We now discuss the issue of precoder design. The general principle for precoding is to have the precoder to be a rate-1 (i.e. no rate loss) recursive convolutional code whose memory does not exceed that of the ISI channel (i.e. no increase of equalizer complexity due to the precoder). Different precoders can make a difference in the performance of the entire coded equalization system. Direct simulations can be used to guide the choice of the precoder, but a more efficient way can make use of the extrinsic information
transfer (EXIT) charts.
%for precoder design and selection, given a PLC channel and an outer code.

%A new precoded turbo equalizer structure with corresponding MAP
%equalizer algorithm for PLC system is proposed in section
%\ref{sec:turbo equalizer}. Besides the outer code and the
%interleaver, the precoder designing is also critical to the turbo
%equalizer performance. In this section, for a given PLC channel and
%outer code, a modified extrinsic information transfer chart (EXIT)
%is proposed for precoder designing and selection

EXIT charts \cite{bib:Brink01} are a powerful tool
for the analysis and evaluation of an iterative decoder by tracking the
evolution of mutual information between the extrinsic information
and the source sequence as the number of iterations increases.
 Its ability to visualize the trajectory of the probabilistic evolution as well as its elegant
properties (such as the area property) %\cite{bib:Ashi04})
make it
extremely popular. It is also employed to predict and analyze the
performance of turbo-like codes to reduce the simulation complexity.

Again, just like the MAP equalizer, the computation of the EXIT curves here must also be tailored to match to the PLC channel characteristics (i.e. precoded ISI channel with non-Gaussian additive noise).

Consider the outer ECC. Let $u\in\{0,1\}$ be an information bit. Let
$L^a_u$ and $L^e_u$ be the input and the output LLR associated with $u$, whose
probabilistic density function (pdf) is given by $p^a_L(L_u)$ and $p^e_L(L_u)$, respectively.
%Without loss of generality, we employ $L_u$ and $p_L(L_u)$ to represent $L^a_u$, $L^e_u$ and $p^a_L(L_u)$, $p^e_L(L_u)$.
For ease of presentation, below we neglect the superscript $a$ and $e$, since the derivations apply to both quantities.

Since the channel is symmetric, we have $p_L(L_u|u=0)=p_L(-L_u|u=0)$. The mutual information between $u$ and $L_u$ can be computed using
%--------------------------------
\begin{align}
    I(u;L_u) =
    &\int\limits^{\infty}\limits_{-\infty}p_L(L_u|u=0)\cdot \nonumber \\
 &
    \log_2 \dfrac{2 p_L(L_u|u\!=\!0)}{(p_L(L_u|u\!=\!0)\!+\!p_L(-L_u|u\!=\!0))}dL_u.
\end{align}
%--------------------------------

Now following the conventional assumption that the message $L_u$ follows a Gaussian distribution
with mean $\mu$ and variance $\sigma^2=2\mu$, the mutual information can
be simplified to
%--------------------------------
\begin{align}\label{equ:mutual information based on Gaussian}
 & I(u;L_u)
   = I_{\mu,\sigma}(\mu,\sigma) \\
   \define & 1
\!-\!\frac{1}{\sqrt{2\pi}\sigma}
    \int_{-\infty}^{\infty}\!\!\!\!\! e^{-(L_u-\mu)^2/2\sigma^2}\log_2(1\!+\!e^{-L_u})dL_u \ \  \mbox{(bit)}.\nonumber
\end{align}
%--------------------------------

\begin{figure}[htf]
\vspace{-0.4cm}
\centerline{
\includegraphics[width=3.8in,height=2in]{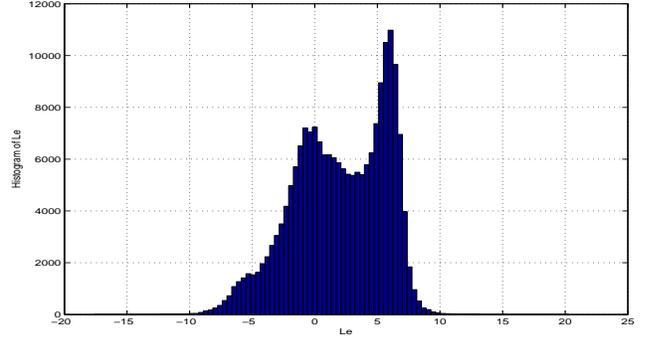}
\vspace{-0.5cm} }
\caption{Histogram of the output LLRs of the MAP
equalizer demonstrating a mixed Gaussian distribution} \label{fig:histogram of Le}
\end{figure}

Similarly, for the inner code (the precoded ISI channel), let $v\in\{0,1\}$ be
an information bit, and $L^a_v$  and $L^e_v$ be the {\it a priori} (input) and the extrinsic (output) LLRs, associated with $v$, whose pdf's are given by $p^a_L(L_u)$ and
$p^e_L(L_u)$ respectively. Since the precoded ISI channel suffers from a non-Gaussian noise, the conventional Gaussian assumption will not apply to the inner code with accuracy. Hence, in stead of Gaussian-based analytical forms, we resort to Monte Carlo simulations. An interesting discovery we made in our simulation study is that, given the mixed Gaussian noise model in (\ref{equ: noise model2}) for PLC, the output LLRs at the MAP equalizer can be modeled by a mixed Gaussian distribution  as shown in Fig. \ref{fig:histogram of Le}.
Since each component of the mixed-Gaussian output LLR corresponds well to the respective component in the mixed-Gaussian noise, following a similar form in (\ref{equ: noise model2}), it is possible to categorize the output extrinsic pdf $p_L(L^e_v)$ by the collection of $D=2$ components, $p_L(L^e_v(i)), \ 0 \!\le\! i\! \le\! D$, each being approximated by a single Gaussian distribution.
The output mutual information from the equalizer can therefore be expressed as
%------------------------
\begin{align}\label{equ:mutual information of inner code}
  &I(v;L^e_v)
  = \sum_{i=0}^{D-1} P^{ini}_i I(v;L^e_v(i)) \nonumber\\
   =& \sum_{i=0}^{D-1} P^{ini}_i
   \int^{\infty}_{-\infty}p_L(L_u|u=0)\cdot \nonumber\\
   & \ \ \ \ \ \  \ \ \log_2 \dfrac{2 p_L(L_u|u=0)}{(p_L(L_u|u=0)+p_L(-L_u|u=0))}dL_u \nonumber \\
   =& \sum_{i=0}^{D-1} P^{ini}_i ( 1
\!-\!\frac{1}{\sqrt{2\pi}\sigma}
    \int_{-\infty}^{\infty}\!\!\!\! e^{-(L_u-\mu)^2/2\sigma^2}\log_2(1\!+\!e^{-L_u})dL_u).\nonumber
\end{align}
%------------------------

%--------------------------------

%---------------------------------
%

%%%%%%%%%%%%%%%%%%%%%%%%%%%%%%%%%%%%%%%%%%%%%%%%%%%%%%%
%%%%%%%%%%%%%%%%%%%%%%%%%%%%%%%%%%%%%%%%%%%%%%%%%%%%%%%
\section{Simulation of Turbo Equalizers} \label{sec:simulation}

\begin{figure}[htbf]
\centerline{
\includegraphics[width=3.8in,height=2in]{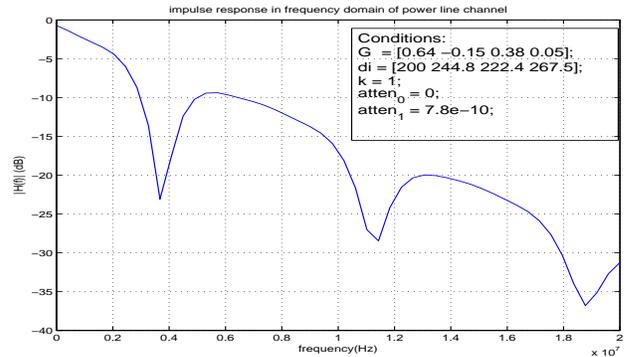}
\vspace{-0.5cm} }
\caption{Frequency impulse response $|H_C(f)|$ of the
PLC channel.} \label{fig:channel frequency response}
\vspace{-0.5cm}
\end{figure}

\begin{figure}[htbf]
\centerline{
\includegraphics[width=3.8in,height=2in]{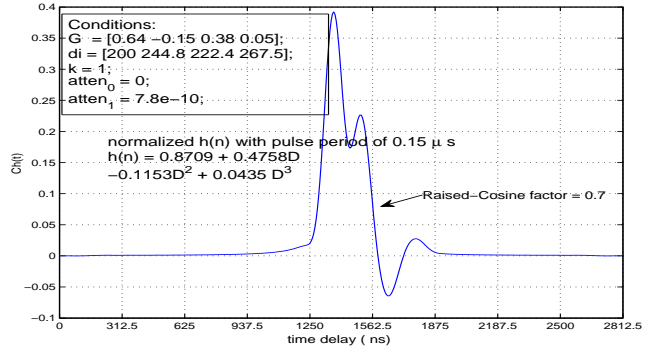}
\vspace{-0.5cm} }
\caption{Impulse response $ch(t)$ of the equivalent
PLC channel with raised-cosine factor of $0.7$ in the time domain.}
\label{fig:channel time response}
\end{figure}

We simulate the proposed precoded turbo equalization scheme on a
$4-path$ power line network model \cite{bib:Zimm02a} with VVF (Vinyl
insulation, Vinyl sheath, Flat) cable ($\varepsilon_r=3.17$). The
channel impulse response in the frequency domain is modeled by
(\ref{equ:frequence channel model}), where the weight factors and
the distances for the $i$th path are $\xi_1 = 0.64$, $\xi_2 =0.38$,
$\xi_3 =-0.15$, $\xi_4 =0.05$, and $d_1 = 200m$, $d_2 = 222.4m$,
$d_3 = 244.8m$, $d_4= 267.5m$. The attenuation factors are $\kappa =
1$, $a_0 = 0$, and $a_1 = 7.8\times 10^{-10} s/m$. The frequency
response of this PLC channel model is demonstrated in Fig.
\ref{fig:channel frequency response}. The pulse is shaped by a
raised-cosine filter with $\beta = 0.7$. The impulse response
$ch(t)$ of the equivalent channel (combining the PLC channel and the
pulse shaping filter) in the time domain is shown in Fig.
\ref{fig:channel time response}. Suppose that an impulse sequence
with a transmission rate of $1/0.15\mu s$ travels along the channel.
We sample the impulse response in the time domain and get a
normalized 4-tap discrete channel model, which is used in all the
simulations shown here:
%------------------------
\begin{equation}\label{equ:discrete channel model}
%    h(t) = 0.1990+0.1267D^{-1}+0.0233D^{-2}+0.0246D^{-3}
   h(t) =  0.8709 +   0.4758 D  -0.1153 D^2 + 0.0435D^3.
\nonumber\end{equation}
%------------------------
%{\bf How about the additive noise? do you use the two-term Gaussian model or the multi-term Gaussian model?}
%--------------------------------
\vspace{-0.4cm}
%---------------------------------
%
%--------------------------------
\vspace{-0.4cm}
%---------------------------------
%
%
%

%%--------------------------------
%\begin{figure}[h]
%\centering
%\includegraphics[width=6in]{pic/fer}
%\caption{The frame error rate of turbo equalizer.} \label{fig:fer}
%\end{figure}
%%---------------------------------

%For reference purpose, we also provide a performance curve assuming
%that the receiver has the perfect side channel information (SCI)
%about when will the pulse noise noise happen. In other words, the
%receiver has a perfect side information about the variance of the
%noise impact on the any symbol. Therefore, the calculation of
%$\gamma$ could be expressed as
%
%\begin{equation}\label{equ: gamma for
%Gaussian noise with SCI}
%    \gamma_k(s_k=s, s_{k-1}=s')  =
%    e^{\frac{1}{2}a_ku_k+\frac{L_c(k)}{2}y_kx_k-\frac{L_c}{4} x_k^2},
%\end{equation}
%where $L_c(k)$ might change for each symbol. Comparing the
%situations between known SCI and unknown SCI, an unexpected
%noticable $2dB$ performance loss is observed from the simulation
%results. This observation motivate us to search for new methjod to
%improve the performance further.

\begin{figure}[htbf]
\centerline{
\includegraphics[width=3.5in,height=2.7in]{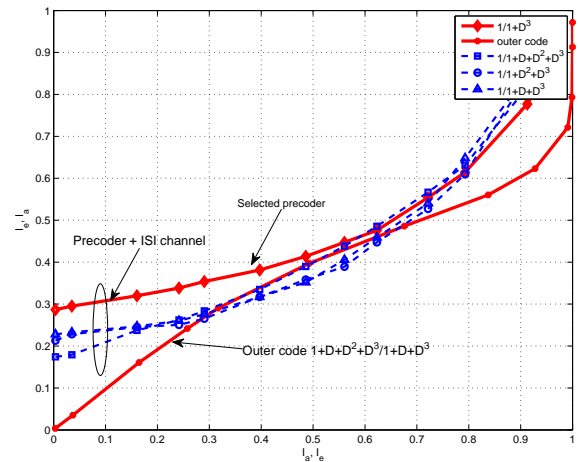}
\vspace{-0.5cm} }
\caption{EXIT charts for the precoded turbo
equalizer. Inner EXIT curve corresponds to precoded PLC channel; outer EXIT curve corresponds to convolutional code [$1, \frac{1+D+D^2+D^3}{1+D+D^3}$].}
\label{fig:exit}
\end{figure}

\begin{figure}[htbf]
\centerline{
\includegraphics[width=3.5in,height=2.6in]{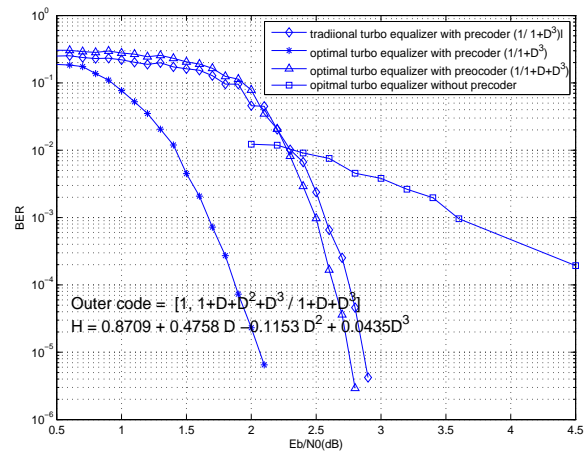}
\vspace{-0.5cm} }
\caption{The bit error rate of turbo equalizer.}
\label{fig:ber}
\end{figure}

We consider the outer ECC code in use is a convolutional code with
generator polynomial [$1, \frac{1+D+D^2+D^3}{1+D+D^3}$]), and look
for a precoder that best matches this ECC code and the PLC channel
in (\ref{equ:discrete channel model}). We do so by studying the EXIT
chart. The best precoder, when applied to the PLC channel, must
exhibit an EXIT curve that matched best, in shape and in position,
with that of the outer ECC. Following the analysis in Section
\ref{sec:exit chart analysis of turbo equalizer}, the EXIT curves
for the precoded PLC channel with four different precoders are
evaluated, and plotted together with the EXIT curve of the outer ECC
code in Figure \ref{fig:exit}. At a signal-to-noise ratio (SNR) of
$-5dB$, we see that most of the precoders have EXIT curves either
touching or crossing that of the ECC code. The only exception is
precoder $\frac{1}{1\oplus D^3}$, whose EXIT curve still leaves a
desirable open tunnel, which will allow the soft information to be
iteratively changed and continually improved without hitting a fixed
point.
% that does not cross with the curve of the outer code
%(i.e. it opens a tunnel for the iteration to proceed through). In
%other words, under the same configuration, the turbo equalizer with
%precoder ($\frac{1}{1+D^3}$) is the only one that can be
%successfully decoded, which will in turn lead to a better
%performance than other precoders.
Hence, we can conclude that precoder $\frac{1}{1\oplus D^3}$ fits
the PLC channel the best. Simulation results of the actual bit error
rate performance  in Fig. \ref{fig:ber} confirms the prediction,
demonstrating a gain of $0.6$dB over other choices of precoders.

To further demonstrate the efficiency of the proposed turbo
equalizer, and especially the importance of the right precoder and
the right MAP algorithm, we compare our performance with two
reference systems in Fig. \ref{fig:ber}. Both reference models are
turbo equalizers. The first uses the optimized precoder
$\frac{1}{1+D^3}$  but the traditional MAP algorithm (not modified
for the PLC channel), and the second reference uses the modified MAP
equalizer discussed in Section \ref{sec:MAP} but no precoders. The
simulation results clearly demonstrate that the  proposed system
gains from both accounts: some $0.7$dB gain from matching the MAP
equalizer to the PLC channel, and more than $2.5$dB gain from
employing a precoder (evaluated at the BER of $10^{-4}$).

%%%%%%%%%%%%%%%%%%%%%%%%%%%%%%%%%%%%%%%%%%%%%%%%%%%%%%%
%%%%%%%%%%%%%%%%%%%%%%%%%%%%%%%%%%%%%%%%%%%%%%%%%%%%%%%
\section{Conclusion}
\label{sec:conclusion}

We have proposed a precoded turbo equalizer scheme for PLC systems
characterized by severe inter-symbol interference and strong pulse
noise. The new scheme transforms the non-recursive ISI channel to
one that is recursive by precoding it with an appropriate rate-1
recursive precoder, hence enabling the renowned interleaving gain.
The new scheme also includes a modified MAP algorithm specifically
designed to address the non-Gaussian pulse noise.
%, which outperforms the
%conventional MAP algorithm (tailored to Gaussian noise).
EXIT charts (with modified algorithm to compute the EXIT curve) are exploited to facilitate the
selection of the right precoder, and extensive simulations
confirm the effectiveness of the proposed scheme.
% we again improve the EXIT chart since the
%noise in PLC channels is non-Gaussian. Our simulation results verify
%that that the new precoded turbo equalizer outperforms the existing
%turbo equalization algorithms in PLC. Simulations also show that
%this dedicated EXIT tool is able to predict precoders performance
%before integrating them in the system and simulation.

% This paper proposed a
%precoded turbo equalizer scheme for power line communication system.
%With a $rate-1$ recursive precoder, the new scheme reforms the
%non-recursive ISI channel into a recursive systematic convolutional
%code, which brings a so-called interleaving gain for a turbo-like
%concatenated system. The corresponding MAP equalizer algorithm in
%the precoded turbo equalizer is optimized for PLC channel to
%mitigate the pulse noise. Given an outer code and discrete channel
%model, a modified EXIT tool is proposed and employed for the design
%and analysis of precoders. The simulation results prove that the new
%precoded turbo equalizer can gain and outperform the existing turbo
%equalizer from three aspects: the precoded scheme, the optimal MAP
%equalizer and the precoder selection.
%

%%--------------------------------
%\begin{figure}[h]
%\centering
%\includegraphics[width=3.5in]{pic/ber_with_diff_precoder}
%\caption{The BER performance of turbo equalizer with different
%equalizer}
%
%\label{fig:ber with different precoder}
%\end{figure}
%%---------------------------------
%\vspace{-0.5cm}

%--------------------------------
\vspace{-0.4cm}
%---------------------------------

%--------------------------------
\vspace{-0.6cm}
%---------------------------------
\end{document}